\documentclass[aps,superscriptaddress,twocolumn,twoside,floatfix,pra,nofootinbib,a4paper]{revtex4-2}
\usepackage{enumerate,appendix}
\usepackage{amsmath, amsthm, amssymb,commath,mathtools}
\usepackage{color,calc,graphicx}
\usepackage[usenames,dvipsnames,svgnames,table,cmyk,hyperref]{xcolor}
\usepackage[colorlinks]{hyperref}
\hypersetup{
	colorlinks = true,
	urlcolor = {blue},
	citecolor = {magenta},
	linkcolor= {blue}
}

\usepackage{graphicx}
\usepackage{amsmath}
\usepackage{latexsym}
\usepackage{bbm}

\newcommand{\ket}[1]{\vert#1\rangle}
\newcommand{\bra}[1]{\langle#1\vert}
\newcommand{\braket}[2]{\langle#1\vert#2\rangle}
\newcommand{\bracket}[3]{\langle#1\vert#2\vert#3\rangle}
\newcommand{\ketbra}[2]{\vert#1\rangle\langle#2\vert}

\usepackage{physics}

\renewcommand{\Tr}[1]{\mathrm{tr}\left[#1\right]}

\usepackage{booktabs}
\usepackage{multirow}
\usepackage{subfigure}
\usepackage{dcolumn}
\usepackage{mathrsfs}
\usepackage{diagbox}
\usepackage{array}
\usepackage{makecell}
\usepackage{threeparttable}

\newtheorem{result}{Result}
\newtheorem{conjecture}{Conjecture}

\usepackage[normalem]{ulem}

\begin{document}


\title{Enhanced Schmidt number criteria based on correlation trace norms}

\author{Armin Tavakoli}
\affiliation{Physics Department, Lund University, Box 118, 22100 Lund, Sweden}

\author{Simon Morelli}
\email{smorelli@bcamath.org}
\affiliation{Basque Center for Applied Mathematics, Mazarredo 14, E48009 Bilbao, Spain}
\affiliation{Atominstitut, Technische Universit\"at Wien, Stadionallee 2, 1020 Vienna, Austria}

\date{\today}

\begin{abstract}
The Schmidt number represents the genuine entanglement dimension of a bipartite quantum state. We derive simple criteria for the Schmidt number of a density matrix in arbitrary local  dimensions, given that certain symmetric measurements exist. They are based on the trace norm of correlations obtained from seminal families of quantum measurements, specifically symmetric informationally complete measurements and  mutually unbiased bases. Our criteria are strictly stronger than both the well-known fidelity witness criterion and the computable cross-norms or realignment criterion. 
\end{abstract}

\maketitle


\section{Introduction}
Entanglement is a paradigmatic resource for quantum information science and technology as well as a fundamental feature of quantum theory. To understand the quantum relation between a pair of particles, it is a natural endeavour to not only ask whether they are entangled but also consider how many of their degrees of freedom genuinely exhibit entanglement. Indeed, if two particles, with one hundred local levels each, are shown to be entangled, we still have no information on how many of these levels are involved in generating the entanglement. This leads to the notion of a genuine entanglement dimension, which is captured by the Schmidt number of a bipartite density matrix. In this more general picture, standard separability corresponds to the special case in which the Schmidt number is just one.  

However, deciding whether a given density matrix is entangled is an NP-hard problem \cite{Gurvits2004, Gharibian2010}, and characterising the Schmidt number appears even harder. Therefore, it is important to derive sufficient conditions for bounding the Schmidt number. Ideally, such a condition should detect many interesting types of entanglement, be valid for any quantum state in arbitrary and unequal  local dimensions, and be easy to compute. Examples of Schmidt number criteria in this spirit are known, for instance by examining the fidelity between the density matrix and the maximally entangled $d$-dimensional state \cite{Terhal2000}.  A different approach is based on generalising both the well-known positive partial-transpose entanglement criterion \cite{Peres1996, Horodecki1996} and the computable cross-norm or realignment criterion (CCNR) \cite{chen2003matrix, Rudolph2005} to apply also for Schmidt numbers \cite{Hulpke2004, johnston2013duality}. A similar criteria based on the Bloch decomposition is known~\cite{Klockl_2015}. Recently, also criteria based on covariance matrices have been developed ~\cite{Liu_2023characterizing,Liu2024bounding}.

In this work, we derive two  simple and general criteria for the Schmidt number. They are each based on analysing the correlations obtained from locally performing certain quantum measurements. These measurements correspond to well-known families, namely symmetric informationally complete (SIC) measurements and  mutually unbiased bases (MUBs). The former corresponds to a maximal set of projections that have identical modulus overlaps \cite{Zauner1999, Renes2004} and the latter corresponds to a maximal set of bases such that any element of any basis when measured in any other basis has uniformly distributed outcomes \cite{Wootters1989}. Both SICs and MUBs have a broad role in quantum information theory. In the specific context of entanglement, they have been used for entanglement witnessing \cite{Spengler2012, Shang2018, Bae2019, Li2022} as well as semi-device-independent \cite{bakhshinezhad2024} and fully device independent entanglement certification \cite{Tavakoli2021}. They can also be used to witness Schmidt numbers via fidelity estimation \cite{Fickler2014, Morelli2023}. 

Our Schmidt number criteria are different; they are based on analysing the trace norm of the correlations obtained when both particles are exposed to either a SIC measurement or MUB measurements. A norm criterion for entanglement detection with SICs was introduced in \cite{Shang2018} and with MUBs in \cite{Siudzinska_2022}, our results can be seen as a generalisation of those to Schmidt number criteria. Importantly, we  prove that both our criteria are strictly stronger than both the fidelity criterion and the CCNR criterion for Schmidt numbers. We further show that in dimensions where both SICs and complete sets of MUBs exists, our two criteria may be closely related. Finally, we show evidence in support of our criteria admitting resource-efficient generalisations, in which fewer measurements are used to detect the Schmidt number. Deciding these conjectures is left as an open problem.


\section{Preliminaries}
We now introduce the necessary preliminaries for the later presentation of our results.


\subsection{Schmidt number}

Consider a bipartite quantum state $\rho_{AB}$ with subsystem dimensions $d_A$ and $d_B$ respectively. The state is said to be entangled if it does not admit a separable decomposition
\begin{equation}\label{sep}
\rho_{AB}= \sum_{\lambda} p_\lambda \phi_\lambda \otimes \varphi_\lambda,
\end{equation}
where $\{p_\lambda\}$ is a probability distribution, and $\phi_\lambda$ and $\varphi_\lambda$ are arbitrary pure states on system $A$ and $B$ respectively. A generalisation of separability is obtained by considering the set of states that have a specific Schmidt number. Specifically, the Schmidt number of $\rho_{AB}$ is the smallest positive integer $r$ for which it is possible to generate the state by stochastically preparing bipartite pure states  that are, up to local unitary transformations, confined to an at most $r$-dimensional local Hilbert space. The latter is equivalent to the Schmidt rank of the pure state, written $\text{SR}(\psi)$, being no more than $r$. The Schmidt rank of a pure state denotes the number of terms appearing in the Schmidt decomposition $ \ket{\psi}=\sum_{i=1}^{\text{SR}(\psi)} \lambda_i \ket{\alpha_i,\beta_i}$, where $\{\ket{\alpha_i}\}_i$ and $\{\ket{\beta_i}\}_i$ are, respectively, orthonormal states and $\{\lambda_i\}_i$ satisfy $\lambda_i> 0$ and $\sum_i \lambda_i^2=1$.
Thus, we may write the Schmidt number as 
\begin{align}\nonumber
r(\rho_{AB})\equiv \min_{\{p_\lambda\},\{\psi_\lambda\}}  \Big\{&r_\text{max}: \quad  \rho_{AB}=\sum_\lambda p_\lambda \ketbra{\psi_\lambda}\\\label{schmidt}
&\text{and} \quad r_\text{max}=\max_\lambda \text{SR}(\psi_\lambda)\Big\}.
\end{align}
Notice that the set of states with Schmidt number $r$ is convex and that in the case of $r=1$ Eq.~\eqref{schmidt} reduces to Eq.~\eqref{sep}.

Let us now discuss two well-known Schmidt number criteria. Consider that we want to determine  the Schmidt number of $\rho_{AB}$ when $d_A=d_B\equiv d$. A sufficient criterion is that the fidelity between $\rho_{AB}$ and the maximally entangled state violates the following inequality \cite{Terhal2000}
\begin{equation}\label{fid}
F(\rho)=\max_{\psi_\text{max}}\hspace{2mm}\bracket{\psi_\text{max}}{\rho_{AB}}{\psi_\text{max}}\leq \frac{r}{d},
\end{equation}
where $\ket{\psi_\text{max}}$ is any maximally entangled state, i.e.~any pure state with maximally mixed reductions. 

A second criterion is the CCNR criterion for Schmidt numbers \cite{johnston2013duality,zhang2023analyzing}.
Let $\{\gamma^A_j\}_{j=1}^{d_A^2}$ and  $\{\gamma^B_j\}_{j=1}^{d_B^2}$ be two orthonormal bases of Hermitian matrices in dimensions $d_A$ and $d_B$ respectively, i.e.~$\tr\left(\gamma_j^A\gamma_k^A\right)=\tr\left(\gamma_j^B\gamma_k^B\right)=\delta_{j,k}$. Then every state $\rho_{AB}$ can be expanded as
\begin{equation}
\rho_{AB}=\sum_{i,j} \tr(\rho_{AB} \gamma_i^A \otimes \gamma_j^B) \gamma_i^A \otimes \gamma_j^B
\end{equation}
and we denote the matrix encoding the coefficients as $[\mathbf{C}]_{ij}=\tr(\rho_{AB} \gamma_i^A \otimes \gamma_j^B)$.
The CCNR criterion stipulates that
\begin{equation}\label{ccnr}
\|\mathbf{C}\|_\text{tr} \leq r,
\end{equation}
for all states with Schmidt number at most $r$, where the trace-norm, also known as the one-norm, is defined as $\|A\|_\text{tr}=\tr\sqrt{A^\dagger A}$ and corresponds to the sum of all singular values of $A$.


\subsection{SICs and MUBs}
In a $d$-dimensional Hilbert space, a SIC-POVM (or SIC measurement)  is defined as a set of positive operators $\{E_i\}_{i=1}^{d^2}$ such that $E_i=\frac{1}{d}\ketbra{\psi_i}{\psi_i}$ and
\begin{equation}
\left|\braket{\psi_j}{\psi_k}\right|^2=\frac{1}{d+1} \quad \forall j\neq k.
\end{equation}
The overlap constant is determined indirectly by the completeness condition of the measurement. SIC-POVMs are known to exist in every dimension up to well-above one hundred \cite{Scott2010, scott2017sics} and are conjectured to exist in every dimension.

A pair of bases of $d$-dimensional Hilbert space are called unbiased if the modulus overlap between any pair of their respective elements is identical. By extension, a set of $m$ bases is called mutually unbiased if every pair of bases are unbiased with respect to each other. Thus, if for each $z=1,\ldots, m$ we have a basis $\{\ket{g_a^z}\}_{a=1}^d$, they are mutually unbiased if and only if 
\begin{equation}\label{MUB}
\left|\braket{g_a^z}{g_{a'}^{z'}}\right|^2=\begin{cases}
\delta_{a,a'} & \text{if } z=z'\\
\frac{1}{d} & \text{if } z\neq z'
\end{cases}.
\end{equation}
For any $d$ there exists at least $m=3$ MUBs and at most $m=d+1$ MUBs. If the upper bound is saturated, the set of MUBs is called a complete set and it is also tomographically complete. Complete sets of MUBs are known to exist in all prime power dimensions \cite{Wootters1989}.


\section{Results}


\subsection{SIC criterion}
Consider that each subsystem of $\rho_{AB}$ is measured with some SIC-POVM, $\{E_a^A\}_{a=1}^{d_A^2}$ and $\{E_b^B\}_{b=1}^{d_B^2}$ respectively. 
The resulting outcome statistics is given by Born's rule, 
\begin{equation}
P^\text{SIC}_{ab}=\tr\left(\rho_{AB} E^A_a \otimes E^B_b\right).
\end{equation}
We write $\mathbf{P}$ for the $d_A^2\times d_B^2$ matrix whose entries are $P^\text{SIC}_{ab}$.  We will now consider the trace norm of $\mathbf{P}$. Our first result shows that this norm acts as a Schmidt number criterion.

\begin{result}[SIC criterion]\label{SICresult}
	For any bipartite state $\rho_{AB}$ of local dimensions $d_A$ and $d_B$ and of Schmidt number at most $r$, it holds that 
	\begin{equation}\label{EAMentcriterion2}
	\|\mathbf{P}\|_\text{tr} \leq \frac{1+r}{K},
	\end{equation}
	where we define a constant depending on the local dimensions $K=\sqrt{d_A(d_A+1)}\sqrt{d_B(d_B+1)}.$
\end{result}

\begin{proof} The index of coincidence for a POVM $\{E_i\}_{i=1}^N$ and state $\sigma$ is defined as $I(\sigma)\equiv \sum_{i=1}^N |\tr(E_i \sigma)|^2$. In the case of $\{E_i\}_{i=1}^N$ being a SIC-POVM, Rastegin derived a bound \cite{Rastegin2021} on $I$ in terms of the purity of $\sigma$. In Appendix~\ref{AppRastegin} we extend the result to the case where $\sigma$ can be any linear operator. The resulting bound is 
	\begin{equation}\label{rasteginbound}
	I(\sigma)\leq \frac{\tr\left(\sigma\right)^2+\tr\left(\sigma^\dagger \sigma\right)}{d(d+1)}.
	\end{equation}
	We will use this as a lemma. Since the trace norm is convex we can w.l.o.g.~restrict our analysis of states with Schmidt number $r$ to pure states with Schmidt rank $r$. Also, due to the freedom of choosing local bases, we can  w.l.o.g.~write the state as $\ket{\psi}=\sum_{s=0}^{r-1} \lambda_s \ket{ss}$, where $\{\lambda_s\}$ is the set of Schmidt coefficients. The probabilities are then given by
\begin{align}\nonumber
P_{ab}&=\sum_{s,t=0}^{r-1} \lambda_s \lambda_t \bracket{ss}{E^A_a\otimes E^B_b}{tt}\\\label{step3}
&=\sum_{s=0}^{r-1}\lambda_s^2 D_s^{a,b} +\sum_{s\neq t} \lambda_s \lambda_tO_{s,t}^{a,b},
\end{align} 
where we have defined $D_s^{a,b}=\bracket{ss}{E^A_a\otimes E^B_b}{ss}$ and $O_{s,t}^{a,b}=\bracket{ss}{E^A_a\otimes E^B_b}{tt}$.  Consider now the trace norm of the matrix $D_s=\sum_{a,b} D_s^{a,b}\ketbra{a}{b}$ and use that $D_s=\ketbra{\alpha_s}{\beta_s}$ where $\ket{\alpha_s}=\sum_a \bracket{s}{E^A_a}{s}\ket{a}$ and $\ket{\beta_s}=\sum_b \bracket{s}{E^B_b}{s}\ket{b}$. A direct calculation gives
\begin{multline}\label{tnorm1}
\|D_s\|_\text{tr}=\sqrt{\braket{\alpha_s}{\alpha_s}} \sqrt{\braket{\beta_s}{\beta_s}} =\\
\sqrt{I_A(\ketbra{s})} \sqrt{I_B(\ketbra{s})}
 \leq \frac{2}{K},
\end{multline}
where we have first used that $\braket{\alpha_r}{\alpha_r}$ and $\braket{\beta_r}{\beta_r}$ each yield the index of coincidence for $\{E_a^A\}$ and $\{E^B_b\}$, and then used the bound \eqref{rasteginbound} for each factor separately. Following the same steps, we can also bound the trace norm of the matrix $O_{s,t}=\sum_{a,b} O_{s,t}^{a,b}\ketbra{a}{b}$ as
\begin{equation}\label{tnorm2}
\|O_{s,t}\|_\text{tr}=\sqrt{I_A(\ketbra{s}{t})}\sqrt{I_B(\ketbra{s}{t})}
\leq \frac{1}{K},
\end{equation}
where we have again used Eq.~\eqref{rasteginbound} but now for the linear operator $\sigma=\ketbra{s}{t}$. 
	
Working from Eq.~\eqref{step3}, we now obtain our final result
\begin{align}\nonumber
	\|\mathbf{P}\|_\text{tr}&\leq\sum_{s=0}^{r-1}\lambda_s^2 \|D_s\|_\text{tr} +\sum_{s\neq t} \lambda_s\lambda_t \|O_{s,t}\|\\\nonumber
	& \leq \frac{1}{K} \left(2\sum_{s=0}^{r-1} \lambda_s^2 + \sum_{s\neq t} \lambda_s\lambda_t\right) \\\nonumber
	& = \frac{1}{K} \left(1  + \bigg(\sum_{s=0}^{r-1} \lambda_s\bigg)^2 \right)\\\nonumber
	& \leq \frac{1+r}{K}. \\
\end{align}
	In the first line we have used the triangle inequality. In the second line we have used the norm bounds  \eqref{tnorm1} and \eqref{tnorm2}. In the third line have we used normalisation $\sum_s \lambda_s^2=1$ and that $\sum_s \lambda_s^2+\sum_{s\neq t}\lambda_s\lambda_t= \sum_{s,t}\lambda_{s}\lambda_t=\left(\sum_s \lambda_s\right)^2$. In the fourth line we have used that Schmidt rank $r$ implies $\left(\sum_s \lambda_s\right)^2\leq r$ \cite{Terhal2000}.
\end{proof}
Note that in the special case of standard separability, namely when $r=1$, the criterion \eqref{EAMentcriterion2} reduces to that introduced in Ref.~\cite{Shang2018}.  It was shown in \cite{Shang2018} that $\|\mathbf{P}\|_\text{tr}$ is invariant under the specific choice of local SIC-POVMs.
To see this, let $\{E_i^{A/B}\}_{i}$ and $\{\Tilde{E}_i^{A/B}\}_{i}$ be two arbitrary normalized SIC POVMs, such that
\begin{align}
[\mathbf{P}]_{ij}&=\tr\left(\rho_{AB} E^A_i \otimes E^B_j\right)\\
[\mathbf{\Tilde{P}}]_{ij}&=\tr\left(\rho_{AB} \Tilde{E}^A_i \otimes \Tilde{E}^B_j\right).
\end{align}
We can now write $\Tilde{E}_i^{A/B}=\sum\limits_{j=1}^{d^2} O^{A/B}_{ij}E^{A/B}_j$ where $O^{A/B}$ are orthogonal matrices, as the angles between SICs remain unchanged. 
More precisely, consider the space of Hermitian $d_{A/B}\times d_{A/B}$ matrices spanned by the set $\{E_i^{A/B}\}_{i}$ with the usual inner product given by the trace of the product of two matrices. Both $\{E_i^{A/B}\}_{i}$ and $\{\Tilde{E}_i^{A/B}\}_{i}$ span the same space and are bases thereof. By definition, the inner product $\tr(E_i^{A/B}E_j^{A/B})=\tr(\Tilde{E}_i^{A/B}\Tilde{E}_j^{A/B})$ remains invariant under the transformation $O^{A/B}$ and therefore the transformation is described by an orthogonal matrix.
It then holds that
\begin{align}
[\mathbf{\Tilde{P}}]_{ij}&=\sum\limits_{k,l=1}^{d^2}O^{A}_{ik}O^{B}_{jl}[\mathbf{P}]_{kl},
\end{align}
or, alternatively, $\mathbf{\Tilde{P}}=O^A\mathbf{P}(O^B)^T$, which implies $\|\mathbf{\Tilde{P}}\|_\text{tr}=\|\mathbf{P}\|_\text{tr}$.


\subsection{MUB criterion}
Consider that subsystem $A$ of $\rho_{AB}$ is measured in the $l$'th basis selected from a complete set of MUBs, $\{\ket{g_a^l}\}_a$. Similarly, subsystem $B$  is measured in the  $k$'th basis of another complete set of MUBs $\{\ket{h_b^k}\}_b$. The statistics is given by  Born's rule
\begin{equation}
Q^\text{MUB}_{al,bk}=\bracket{g_a^l,h_b^k}{\rho_{AB}}{g_a^l,h_b^k}.
\end{equation}
We associate this $d_A(d_A+1)\times d_B(d_B+1)$ table of probabilities to the matrix $\mathbf{Q}=\sum_{a,l,b,k}Q^\text{MUB}_{al,bk} \ketbra{a,l}{b,k}$. Our second main results is a Schmidt number criterion based on the trace norm of this data table. 

\begin{result}[MUB criterion]\label{MUBresult}
	For any bipartite state $\rho_{AB}$ of local dimensions $d_A$ and $d_B$ and of Schmidt number at most $r$, it holds that 
	\begin{align}\label{MUBequation}
	\left\|\mathbf{Q}\right\|_{tr}\leq 1+r.
	\end{align}
\end{result}

\begin{proof}
For an arbitrary positive semidefinite operator $\sigma$, it follows from the uncertainty relations derived in \cite{Larsen1990, Ivanovic1992, Wu2009} that 
\begin{align}\label{lemma}
	\sum\limits_{l=1}^{d+1}\sum\limits_{a=0}^{d-1}|\tr\left(G_a^l\sigma \right)|^2&\le \tr(\sigma^\dagger \sigma)+\tr(\sigma)^2,
\end{align}
for $G_a^l=\ketbra{g_a^l}{g_a^l}$ (and similarly $H_b^k=\ketbra{h_b^k}{h_b^k}$). 
	We will use this as a lemma.

By convexity of the trace norm, we need only to consider pure states of Schmidt rank $r$. Without loss of generality, we can represent the state in the computational basis, $	|\psi\rangle=\sum\limits_{s=0}^{r-1}\lambda_s|s,s\rangle$.  We can then write
	\begin{multline}
	\left[\mathbf{Q}\right]_{al,bk}=\sum\limits_{s,t=0}^{r-1}\lambda_s\lambda_t \tr\left(\ketbra{t}{s} G_a^l\right)\tr\left(\ketbra{t}{s} H_b^k\right).
	\end{multline}
Define  $[D_s]_{al,bk}= \tr\left(\ketbra{s}{s} G_a^l\right)\tr\left(\ketbra{s}{s} H_b^k\right)$ and   $[O_{s,t}]_{al,bk}=\tr\left(\ketbra{t}{s} G_a^l\right)\tr\left(\ketbra{t}{s} H_b^k\right)$. It then follows that	
	\begin{align}\nonumber
	&\|D_s\|_\text{tr}=\sqrt{\sum_{l=1}^{d+1}\sum_{a=0}^{d-1} \bracket{s}{G_a^l}{s}^2}\sqrt{\sum_{k=1}^{d+1}\sum_{b=0}^{d-1} \bracket{s}{H_b^k}{s}^2}\\\nonumber
	&\|O_{s,t}\|_\text{tr}=\sqrt{\sum_{l=1}^{d+1}\sum_{a=0}^{d-1} |\bracket{s}{G_a^l}{t}|^2}\sqrt{\sum_{k=1}^{d+1}\sum_{b=0}^{d-1} |\bracket{s}{H_b^k}{t}|^2}.
	\end{align}
	Applying Eq.~\eqref{lemma} to each factor above, we obtain that $\|D_s\|_\text{tr}\leq 2$ and $\|O_{s,t}\|_\text{tr}\leq 1$. Using these bounds and basic properties of Schmidt coefficients, we arrive at 
	\begin{align}\nonumber
	\left\|\mathbf{Q}\right \|_\text{tr}&\leq \sum\limits_{s=0}^{r-1}\lambda_s^2 \|D_s\|_\text{tr}+\sum\limits_{s\neq t}\lambda_s\lambda_t \|O_{s,t}\|_\text{tr}\\\nonumber
	&\leq 2\sum\limits_{s=0}^{r-1}\lambda_s^2+\sum\limits_{s\neq t}\lambda_s\lambda_t\\
	& = 1+ \left(\sum_{s=0}^{r-1} \lambda_s\right)^2\leq 1+r.
	\end{align}
\end{proof}

Note that in the special case of standard separability, namely when $r=1$, the criterion \eqref{MUBequation} reduces to that introduced in Ref.~\cite{Siudzinska_2022}. 
Similar to the SIC case, also the quantity $\|\mathbf{Q}\|_\text{tr}$ is invariant under the specific choice of local MUBs. The reasoning follows the same lines, with the only difference that the projectors of the MUBs form a spanning set instead of a basis of the space of Hermitian matrices.
Assume there exist two inequivalent complete sets of MUBs in system $A$ and $B$ respectively. Let $\{G_a^l\}_{a,l}$ and $\{\Tilde{G}_a^l\}_{a,l}$ be the respective basis elements of these two sets in $A$ and $\{H_b^k\}_{b,k}$ and $\{\Tilde{H}_b^k\}_{b,k}$ the respective basis elements of the two complete sets in $B$. It now holds that
\begin{align}
[\mathbf{Q}]_{al,bk}&=\tr\left(\rho_{AB} G^l_a \otimes H^k_b\right)\\
[\mathbf{\Tilde{Q}}]_{al,bk}&=\tr\left(\rho_{AB} \Tilde{G}^l_a \otimes \Tilde{H}^k_b\right).
\end{align}
Since the set $\{G_a^l\}_{a,l}$ spans the space of Hermitian $d_A\times d_A$ matrices, we can write $\Tilde{G}_a^l=\sum\limits_{a',l} O^{A}_{al,a'l'}G^{l'}_{a'}$. The same holds for $\{H_b^k\}_{b,k}$ in system $B$, $\Tilde{H}_b^k=\sum\limits_{b',k'} O^{B}_{bk,b'k'}H^{k'}_{b'}$.
Since the operator $O^{A}$ preserves the angles between the spanning sets $\{G_a^l\}_{a,l}$ and $\{\Tilde{G}_a^l\}_{a,l}$, $O^A$ is an orthogonal matrix. The same holds for $O^{B}$.
It then holds that
\begin{align}
[\mathbf{\Tilde{Q}}]_{al,bk}&=\sum\limits_{a',l',b',k'}O^{A}_{al,a'l'}O^{B}_{bk,b'k'}[\mathbf{Q}]_{a'l',b'k'},
\end{align}
or, alternatively, $\mathbf{\Tilde{Q}}=O^A\mathbf{Q}(O^B)^T$, which implies $\|\mathbf{\Tilde{Q}}\|_\text{tr}=\|\mathbf{Q}\|_\text{tr}$.


\subsection{Relation to fidelity criterion and CCNR criterion}
In the special case in which the local dimensions are equal, we can compare the criteria in Results~\ref{SICresult} and~\ref{MUBresult} to the fidelity criterion \eqref{fid}. We now show that the formers are strictly stronger than the latter. 

To this end, let us begin with rederiving the fidelity criterion \eqref{fid} from Result~\ref{SICresult}. Note that the maximally entangled state $\ket{\phi^+}=\frac{1}{\sqrt{d}}\sum_{i=0}^{d-1}\ket{ii}$ can be written as \cite{Morelli2023}
\begin{equation}
\ketbra{\phi^+}=(d+1)\sum_{i=1}^{d^2} E_i \otimes E_i^*-\frac{1}{d}\openone,
\end{equation}
where $\{E_i\}$ is any SIC-POVM. Since any maximally entangled state takes the form $\ket{\psi_\text{max}}=\openone\otimes U \ket{\phi^+}$ for some unitary, we can define the pair of SIC-POVMs $E^A_i=E_i$ and $E^B_i=U(E_i)^* U^\dagger$ and write 
\begin{equation}
\ketbra{\psi_\text{max}}=(d+1)\sum_{i=1}^{d^2} E_i^A \otimes E_i^B-\frac{1}{d}\openone.
\end{equation}
For the optimal choice of $\psi_\text{max}$, the fidelity becomes
\begin{equation}
F(\rho)=(d+1)\sum_i\tr(\rho E_i^A\otimes E_i^B)-\frac{1}{d}=(d+1)\sum_iP_{ii}^\text{SIC}-\frac{1}{d}.
\end{equation}
Note that $\sum_i P_{ii}^\text{SIC}=\tr(\mathbf{P})$. It then holds that
\begin{equation}
\tr(\mathbf{P})=\sum\limits_i\lambda_i(\mathbf{P})\le\sum\limits_i |\lambda_i(\mathbf{P})|\le\sum\limits_i \sigma_i(\mathbf{P})=\left\|\mathbf{P}\right \|_\text{tr},
\end{equation}
where we have used that the sum of the modulus eigenvalues of any square matrix is bounded by the sum of its singular values $\sigma_i$. In conclusion,
\begin{equation}\label{fidelitySIC}
F(\rho)\leq (d+1)\|\mathbf{P}\|_\text{tr}-\frac{1}{d}\leq \frac{r}{d},
\end{equation}
where in the last step we used the Schmidt number bound from Result~\ref{SICresult}. Thus we have recovered the fidelity criterion \eqref{fid}.

In a similar way, the fidelity criterion can also be recovered from Result~\ref{MUBresult}. One notes that for any complete set of MUBs \cite{Morelli2023}
\begin{equation}
\ketbra{\phi^+}{\phi^+}=\frac{1}{d}\sum_{z=1}^{d+1}\sum_{a=1}^d G_a^z\otimes (G_a^z)^*-\frac{1}{d}\openone,
\end{equation}
where $G_a^z=\ketbra{g_a^z}{g_a^z}$.
Defining $H^z_a=U(G_a^z)^* U^\dagger$ we write 
\begin{equation}
\ketbra{\psi_\text{max}}=\frac{1}{d}\sum_{z=1}^{d+1}\sum_{a=1}^{d} G_a^z \otimes H_a^z-\frac{1}{d}\openone.
\end{equation}
For the optimal choice of $\psi_\text{max}$, the fidelity becomes
\begin{equation}
F(\rho)=\frac{1}{d}\sum_{a,z}\tr(\rho G_a^z\otimes H_a^z)-\frac{1}{d}=\frac{1}{d}\sum_{a,z}Q_{az,az}^\text{MUB}-\frac{1}{d}.
\end{equation}
Noting that $\sum_{a,z}Q_{az,az}^\text{MUB}=\tr(\mathbf{Q})\le\left\|\mathbf{Q}\right \|_\text{tr}$ we
finally find
\begin{equation}\label{fidelityMUB}
F(\rho)\leq \frac{1}{d}\|\mathbf{Q}\|_\text{tr}-\frac{1}{d}\leq \frac{r}{d},
\end{equation}
where we used the Schmidt number bound from Result~\ref{MUBresult} and again recover the fidelity criterion \eqref{fid}.

When the SIC criterion for entanglement detection was introduced in Ref.~\cite{Shang2018}, it was conjectured to be strictly stronger than the CCNR criterion \eqref{ccnr}.
This conjecture was stated in the form that $\left\|\mathbf{C}\right \|_\text{tr}>1$ implies $\left\|\mathbf{P}\right \|_\text{tr}>2/K$.
This conjecture was later proven correct in Ref.~\cite{jivulescu2020multipartite}, by showing that $\left\|\mathbf{P}\right \|_\text{tr}\ge (\left\|\mathbf{C}\right \|_\text{tr}+1)/K$.
This result immediately implies that also the SIC criterion for Schmidt number detection is strictly stronger than the CCNR criterion for Schmidt numbers.
To witness Schmidt number $r+1$ with the CCNR criterion, one needs to observe $\left\|\mathbf{C}\right \|_\text{tr}> r$. This in turn implies $\left\|\mathbf{P}\right \|_\text{tr}>(r+1)/K$, meaning that the SIC criterion witnesses the same Schmidt number $r+1$.

In Ref.~\cite{Shang2018} examples of states are given, where entanglement is detected by the SIC criterion, but not by the CCNR or PPT criterion. We now present an example where our criteria outperform both the fidelity criterion and the CCNR criterion for the detection of Schmidt number 3. Define the state $\rho(q)=q\ketbra{\phi^+}+(1-q)\ket{0}\bra{1}$ in dimension 3. It clearly hods that $F(\rho)=q$ and therefore Schmidt number 3 is detected only for $q>2/3$ by the fidelity criterion. The CCNR criterion detects Schmidt numer 3 for $q>0.6162$, whereas both our criteria do so already for $q>0.6089$. This is striking, as the improvement in the detection is already visible for a simple family of states in dimension 3. We expect this improvement to be more visible for specifically designed families of states and also increase with the dimension.


\subsection{Towards equivalence of the criteria}
It is natural to ask whether there is a relation between Result~\ref{SICresult} and Result~\ref{MUBresult}. To study this, we use that Ref.~\cite{Beneduci_2013} introduces a way to constructively obtain a SIC-POVM from a complete set of MUBs. This method gives a set of operators that form a SIC-POVM if and only if the operators are positive, which cannot be a priori guaranteed. We will use this result to show the following.

\begin{result}\label{equivalence}
Assume that a complete set of MUBs exists in both dimensions $d_A$ and $d_B$, and apply to these the construction of Theorem 3 of Ref.~\cite{Beneduci_2013}. If this construction results in a valid SIC-POVM, the criteria of Result~\ref{SICresult} and~\ref{MUBresult} are equivalent.
\end{result}

\begin{proof}
Let $\{G_i\}_{i=1}^{d(d+1)}$ be a complete set of MUB projectors and $\{E_i\}_{i=1}^{d^2}$ a set of SIC projectors as defined above. Since both of these are spanning sets for the Hermitian $d\times d$ operators, there must exist a $d^2\times d(d+1)$ operator mapping one set to the other.
If it succeeds, the construction of Ref.~\cite{Beneduci_2013} explicitely gives such an operator $\mathbf{\Theta}$.
It holds
\begin{align}
	E_i=\sum\limits_{j=1}^{d(d+1)}\mathbf{\Theta}_{ij}G_j=\frac{1}{d\sqrt{d+1}}\sum\limits_{j\in I_i}^{d(d+1)}G_j+\frac{1-\sqrt{d+1}}{d^2}\openone,
\end{align}
where the $d^2$ sets $I_i$ are chosen by taking one element of each MUB in such a way that any two sets share exactly one element. The identity matrix can be written as the sum of the elements of any basis. Splitting the identity evenly as a sum of the $(d+1)$ bases, we arrive at
\begin{align}
	\mathbf{\Theta}=\frac{1}{d\sqrt{d+1}}M+\frac{1-\sqrt{d+1}}{d^2(d+1)}C,
\end{align}
where all entries in $M$ are either 0 or 1, such that each row has exactly $d+1$ entries 1 and any two rows have exactly one shared entry 1. All entries in $C$ are 1.
Using that $MM^T=d\openone+C$, $MC^T=CM^T=(d+1)C$ and $CC^T=d(d+1)C$, it follows
$\mathbf{\Theta}\mathbf{\Theta}^T=\openone/(d(d+1))$. Therefore the singular value decomposition is $\mathbf{\Theta}= O_1 \Delta O_2^T$ with
\begin{equation}
	\Delta = \begin{pmatrix}\tfrac{1}{\sqrt{d(d+1)}} &&& 0 & \hdots &0\\
    & \ddots && \vdots & \ddots & \vdots\\
    && \tfrac{1}{\sqrt{d(d+1)}} & 0 & \hdots &0
    \end{pmatrix}.
\end{equation}

It now holds that
\begin{align}
	[\mathbf{P}]_{ij}&=\Tr{\rho E_i^{A}\otimes E_j^{B}}\\
	&=\Tr{\rho\ (\sum\limits_{k}\mathbf{\Theta}_{ik}^{A}G_k^{A})\otimes (\sum\limits_{l}\mathbf{\Theta}_{jl}^{B}G_l^{B})}\\
	&= \sum\limits_{k,l}\mathbf{\Theta}_{ik}^{A}\mathbf{\Theta}_{jl}^{B}\Tr{\rho\ G_k^{A}\otimes G_l^{B}}\\
	&= \sum\limits_{k,l}\mathbf{\Theta}_{ik}^{A}\mathbf{\Theta}_{jl}^{B}[\mathbf{Q}]_{kl}
\end{align}
and therefore $\mathbf{P}=\mathbf{\Theta}^{A}\mathbf{Q}(\mathbf{\Theta}^{B})^{T}$.

It follows that
\begin{align}
	\left\|\mathbf{P}\right \|_\text{tr}&=\left\|\mathbf{\Theta}^{A}\mathbf{Q}(\mathbf{\Theta}^{B})^{T}\right\|_\text{tr}=\left\|\Delta^{A}\right\|_2 \left\|\Delta^{B}\right\|_2 \left\|\mathbf{Q}\right\|_\text{tr}\\
	&=\frac{1}{K}\left\|\mathbf{Q}\right\|_\text{tr},\notag
\end{align}
where $K$ is defined as in Eq.~\eqref{SICresult}. Equality in the second step comes from the fact that all singular values of $\Delta^{A/B}$ are equal and that $\|\mathbf{Q}\|$ can have at most $\min(d_A^2,d_B^2)$ singular values.
The claim follows immediately.

\end{proof}

Based on this result, we put forward the following conjecture.
\begin{conjecture}\label{equivalence}
Provided that both a SIC and a complete set of MUBs exist in dimension $d_A$ and $d_B$, the criteria in Result~\ref{SICresult} and~\ref{MUBresult} are equivalent. It then holds that $\left\|\mathbf{Q}\right\|_\text{tr}=K\left\|\mathbf{P}\right \|_\text{tr}$.
\end{conjecture}
Explicit calculations show that this is indeed the case for prime dimensions smaller than 20.
These findings bear the question if the quantity $\left\|\mathbf{Q}\right\|_\text{tr}=K\left\|\mathbf{P}\right \|_\text{tr}$ can be used to define an entanglement monotone. In Appendix~\ref{Appmonotone} we show that this is not the case by explicitely constructing a counterexample.


\section{Examples: depolarising and dephasing noise}
We consider two commonly used noise models that naturally occur in many quantum information scenarios, namely when a maximally entangled state is subject to noise of uniform spectral density (depolarisation) and noise causing decoherence in a given basis (dephasing). A depolarisation of the maximally entangled state leads to an isotropic state, which takes the form
\begin{equation}
\rho_v^\text{iso}=v\ketbra{\phi^+_d} +\frac{1-v}{d^2}\openone,
\end{equation}
where $\ket{\phi^+_d}=\frac{1}{\sqrt{d}}\sum_{i=0}^{d-1}\ket{i,i}$ is the maximally entangled state and $v\in[0,1]$ is the visibility of the noisy state. It is known that for the isotropic state to have Schmidt number $r+1$ the visibility $v$ must be greater than the critical value $v_\text{opt}=\frac{rd-1}{d^2-1}$ \cite{Terhal2000}.

In a similar vein, a dephasing in the computational basis of the maximally entangled state leads to the noisy state
\begin{equation}
\rho_u^\text{deph}=u \ketbra{\phi^+_d}+\frac{1-u}{d}\sum_{i=0}^{d-1}\ketbra{i,i}{i,i},
\end{equation}
for some visibility $u\in[0,1]$. The critical visibility for the dephased state having Schmidt number $r+1$ is known to be $u_\text{opt}=\frac{r-1}{d-1}$ \cite{Bavaresco2018}.

We shall now see that both the SIC-criterion and the MUB-criterion for high-dimensional entanglement detection can exactly recover these optimal visibility thresholds. 
For any value of $v>v_\text{opt}$ and $u>u_\text{opt}$ it holds that
\begin{align}
	F(\rho_v^\text{iso})=\frac{1+(d^2-1)v}{d^2}>\frac{r}{d}\\
	F(\rho_u^\text{deph})=\frac{1+(d-1)u}{d}>\frac{r}{d}
\end{align}
and so the fidelity witness is optimal in both cases.
We have already shown that our criteria are strictly stronger than the fidelity witness. In fact it follows immediately from Eqs.~\eqref{fidelitySIC} and~\eqref{fidelityMUB} that
\begin{align}
\|\mathbf{P}\|_\text{tr}&\ge \frac{dF(\rho)+1}{d(d+1)}>\frac{r+1}{d(d+1)},\\
\|\mathbf{Q}\|_\text{tr}&\ge dF(\rho)+1>r+1
\end{align}
and therefore also the SIC and MUB criterion are optimal.


\section{Conjectures}

In this section we investigate the case of more general sets of measurements and put forward two conjectures concerning their potential use for Schmidt number detection.


\subsection{Equiangular measurements}
We now extend Result~\ref{SICresult} to more general sets of symmetric measurements that are not necessarily tomographically complete.
Consider a set of $n>d$ pure states $\{\ket{\psi_a}\}_{a=1}^n$ in $d$-dimensional Hilbert space. We call the set an equiangular measurement (EAM) if the following two conditions are satisfied. Firstly, by defining the subnormalised projector $E_a=d/n\ketbra{\psi_a}$, we can interpret $\{E_a\}_a$ as a quantum measurement. This is equivalent to requiring that the quantum states resolve the identity.
\begin{equation}\label{EAM1}
\sum_{a=1}^n \ketbra{\psi_a}=\frac{n}{d}\openone_d.
\end{equation}
Secondly, the quantum states are equiangular, i.e.~the modulus overlap of all distinct states ($a\neq a'$) is constant;
\begin{equation}\label{EAM2}
\left|\braket{\psi_a}{\psi_{a'}}\right|^2=\frac{n-d}{d(n-1)}.
\end{equation}
In the case of $n=d^2$ we recover the definition of a SIC-POVM.

In analogy with the approach leading up to Result~\ref{SICresult},  we can define the $n\times n$ matrix $\|\mathbf{P}_{n}\|_\text{tr}$ encoding the outcome statistics when measuring with two EAMs $\{E_a^A\}_a^{n}$ and $\{E_b^B\}_b^{n}$. Then for any state of Schmidt number $r$ it holds that
\begin{align}\label{EAMbound}
    \|\mathbf{P}_{n}\|_\text{tr}\le \frac{n-d+d(d-1)r}{n(n-1)}.
\end{align}
In contrast to the full SIC case, the quantity $\|\mathbf{P}_n\|_\text{tr}$ depends on the specific choice of EAM.
The proof of this statement is given in Appendix~\ref{AppEAMproof}. Note that we assume here both the dimension and the number of elements in the EAMs to be equal for both subsystems. The proof for the general case of different local dimensions and number of EAM elements works analogously.

The bound in Eq.~\eqref{EAMbound} is not tight, except when $n=d^2$ or $r=1$. For the latter we recover the result proven in~\cite{Shi2024}. We conjecture the following refinement of the statement.

\begin{conjecture}
For any state of Schmidt number $r$ it holds that
\begin{align}
    \|\mathbf{P}_{n}\|_\text{tr}\le \frac{d(d-1)+r(n-d)}{n(n-1)}.
\end{align}
\end{conjecture}
For $n=d^2$ or $r=1$ this coicides with Eq.~\eqref{EAMbound}.


\subsection{Incomplete sets of MUBs}

If only a subset of $m<d+1$ MUBs is used, we can define the $md\times md$ matrix $\|\mathbf{Q}_{m}\|_\text{tr}$ encoding the outcome statistics when measuring with $m d$ local MUB projectors $\{E_a^l\}_{a,l}$ and $\{E_b^k\}_{b,k}$. Result~\ref{MUBresult} can immediately be extended to 
\begin{align}\label{ncMUBbound}
    \|\mathbf{Q}_m\|_\text{tr}\le \frac{m-1}{d} + r,
\end{align}
where $\mathbf{Q}_m$ encodes the probabilities of $m$ MUBs. In contrast to the complete MUB case, the quantity $\|\mathbf{Q}_m\|_\text{tr}$ depends on the set of chosen MUBs.
Take for example the state $\ket{\phi}=(\ket{00}+\ket{11})/\sqrt{2}$ in dimension 3. When measuring in the standard basis and any other unbiased basis of the Wootters-Fields construction~\cite{Wootters1989}, we obtain $\|\mathbf{Q}_2\|_\text{tr}=5/3$ and thus detect entanglement. Measuring any two unbiased bases of the Wootters-Fields construction except the standard basis results in  $\|\mathbf{Q}_2\|_\text{tr}=4/3$ and no entanglement is detected.
As for EAMs, we assume the local dimension and number of MUBs to be equal for system A and B, but also the general case follows straightforewardly from the proof of Result~\ref{MUBresult}.
The bound given in Eq.~\eqref{ncMUBbound} is generally not tight for $m<d+1$ and we conjecture the following refinement.

\begin{conjecture}
For any state of Schmidt number $r$ it holds that
\begin{align}
    \|\mathbf{Q}_m\|_\text{tr}\le 1 + \frac{(m-1)r}{d}.
\end{align}
\end{conjecture}
For $m=d+1$ or $r=1$ this coincides with Eq.~\eqref{ncMUBbound}.


\section{Discussion}
We have presented two criteria for Schmidt number detection in arbitrary-dimensional systems. These criteria make use of the table of measurement statistic obtained when measuring in the set of projectors forming a symmetric informationally complete POVM or a complete set of mutually unbiased bases. Both criteria are strictly stronger than the fidelity criterion and the CCNR criterion. In fact, it is not to hard to construct examples of states which are detected by our criteria but not by the two others. Under certain circumstances, we have found that our two criteria can be shown to be equivalent and we conjecture this to always be the case, insofar both constructions exist in a given dimension. This connection between the SIC-based and the MUB-based criterion does not leave the other redundant, but contrarily gives two options to test the same criterion. Indeed, both the existence and construction of SIC-POVMs and complete sets of MUBs are not known in arbitrary-dimensions. Here the two criteria complement each other. While the SICs up to some low three digit dimensions can efficiently be obtained by the results of Ref.~\cite{Scott2010, scott2017sics}, MUBs can be constructed in arbitrary prime power dimension~\cite{Wootters1989}.

The motivation for the presented criteria is not experimental applicability but primarily a theoretical understanding of high-dimensional entanglement. Nonetheless, they can still be relevant for determining Schmidt numbers in experiments. Measuring a complete set of MUBs or SIC projectors is experimentally as complex as full state tomography and the state can be  reconstructed by the obtained statistics. However, it is not necessary to measure all projectors forming a SIC or a complete set of MUBs to apply our criteria. The singular values of any diagonal submatrix are smaller than the singular values of the full correlation matrix. It is hence possible to identify subsets of projectors that already lead to a violation of the presented bounds, without the need to fully reconstruct the correlation matrix $ \mathbf{Q}$ or $\mathbf{P}$. Additionally, our results can be adopted in designing experiments, as they bound the Schmidt number that can be experimentally witnessed with a fidelity witness.

Our work leaves several conjecture that are interesting to decide. Another relevant question is whether also other norms can be used to obtain (perhaps even stronger) Schmidt number criteria. Finally, it may also be  interesting to consider whether norm-based entanglement criteria can be extended from our bipartite consideration to also detect the dimension of genuine multipartite entanglement \cite{Cobucci2024}.


\begin{acknowledgements}
The authors are thankful to Shuheng Liu, Jens Siewert, Otfried G\"uhne, Ion Nechita, Robin Krebs and Mariami Gachechiladze for useful discussions. 
A.T.~ was supported by the Wenner-Gren Foundation, by the Swedish Research Council under Contract No.~2023-03498 and by the Knut and Alice Wallenberg Foundation through the Wallenberg Center for Quantum Technology (WACQT).
S.M. was supported by the Basque Government through IKUR strategy and through the BERC 2022-2025 program and by the Ministry of Science and Innovation: BCAM Severo Ochoa accreditation CEX2021-001142-S / MICIN / AEI / 10.13039/501100011033, PID2020-112948GB-I00 funded by MCIN/AEI/10.13039/501100011033 and by "ERDF A way of making Europe".
\end{acknowledgements}

\bibliography{SN_mubsic_references}

\appendix

\section{Extending Rastgin's proof to general operators}\label{AppRastegin}

In this appendix we give a proof that the index of coincidence $I(\sigma)\equiv \sum_{i=1}^{d^2} |\tr(E_i \sigma)|^2$ for a SIC-POVM $\{E_i\}_{i=1}^{d^2}$ and any linear operator $\sigma$ satisfies the relation

\begin{align}
    I(\sigma)=&\frac{\tr(\sigma)^2+\tr(\sigma^\dagger\sigma)}{d(d+1)}.
\end{align}

Let $\{E_i\}_{i=1}^{d^2}$ be a SIC-POVM and define
\begin{align}
    F_i=\sqrt{d(d+1)}E_i-\frac{\sqrt{d+1}-1}{d\sqrt{d}}\openone.
\end{align}
A quick calculation shows that $\tr(F_iF_j)=\delta_{ij}$ and so $\{F_i\}_{i=1}^{d^2}$ forms an orthonormal basis for the linear operators. Therefore for any operator $\sigma$ it holds that
\begin{align}
    \sigma&=\sum\limits_{i=1}^{d^2}\tr(F_i\sigma)F_i\\
    &=\sum\limits_{i=1}^{d^2}d(d+1)\tr(E_i\sigma)E_i-\tr(\sigma)\openone.
\end{align}
We subtract $\tr(\sigma)\openone$, multiply each side with $\sigma^\dagger+\tr(\sigma)\openone$ and take the trace, resulting in
\begin{align}
    &\tr(\sigma^\dagger \sigma)+(d+2)\tr(\sigma)^2\\
    &=\tr(\sum\limits_{i,j=1}^{d^2}d^2(d-1)^2\tr(E_i\sigma)^*\tr(E_j\sigma)E_iE_j).
\end{align}
Using $\tr(E_iE_j)=\delta_{ij}/(d(d+1))+1/(d^2(d+1))$ and $\sum\limits_{i,j=1}^{d^2}\tr(E_i\sigma)^*\tr(E_j\sigma)=\tr(\sigma)^2$, we conclude
\begin{align}
    I(\sigma)=\frac{\tr(\sigma)^2+\tr(\sigma^\dagger \sigma)}{d(d+1)}.
\end{align}

In Appendix~\ref{AppEAMproof} we extend this proof to general equiangular tight frames (EAMs).


\section{Entanglement monotones}\label{Appmonotone}

We have seen that both Result~\ref{SICresult} and~\ref{MUBresult} are invariant under local unitaries, more efficient than the CCNR criterion and conjectured to be ultimately equivalent. These facts bear the obvious question if these quantities can be used to define entanglement monotones of the form $f(\|\mathbf{P}(\rho)\|_{\text{tr}})$, where $f:\mathbb{R}\rightarrow\mathbb{R}$ is a monotonically increasing and convex function.

It clearly holds that the matrix $\mathbf{P}$ is additive
\begin{align}
    \mathbf{P}(\lambda\rho+(1-\lambda)\sigma)=\lambda \mathbf{P}(\rho)+(1-\lambda) \mathbf{P}(\sigma)
\end{align}
and from the convexity of $f$ and the triangle-inequality of the trace norm it holds that
\begin{align}
    &f(\|\mathbf{P}(\lambda\rho+(1-\lambda)\sigma)\|_\text{tr})\\
    &\le \lambda f(\|\mathbf{P}(\rho)\|_\text{tr})+(1-\lambda)f(\|\mathbf{P}(\sigma)\|_\text{tr}).
\end{align}

According to~\cite{Vidal_2000}, in order to show that $f(\|\mathbf{P}(\rho)\|_{\text{tr}})$ is an entanglement monotone, one has to prove that 
\begin{align}
    f(\|\mathbf{P}(\rho)\|_\text{tr})\ge \sum\limits_k p_k f(\|\mathbf{P}(\rho_k)\|_\text{tr}),
\end{align}
where $p_k=\tr{\mathcal{E}_k(\rho)}$ and $\rho_k=\mathcal{E}_k(\rho)/p_k$ for any unilocal quantum operation $\mathcal{E}_k$ performed by any party (local instrument).

However, it is possible to increase the correlation norms by local operations.
Take for example the state $\rho_{ABA'B'}=\ketbra{\phi^+}\otimes\openone/2\otimes\openone/2$. The CCNR criterion and the SIC-criterion then become 
\begin{align}
	&\left\|\mathbf{C}(\rho_{ABA'B'})\right \|_{\text{tr}}=1\\
    K&\left\|\mathbf{P}(\rho_{ABA'B'})\right \|_{\text{tr}}=2.
\end{align}
Two projective measurements in $\sigma_Z$ in system $A'$ and $B'$ result either in $\rho'_{ABA'B'}=\ketbra{\phi^+}\otimes\ketbra{i}\otimes\ketbra{i}$ or in $\rho''_{ABA'B'}=\ketbra{\phi^+}\otimes\ketbra{i}\otimes\ketbra{j}$.
It holds that
\begin{align}
    &\left\|\mathbf{C}(\rho'_{ABA'B'})\right \|_{\text{tr}}=\left\|\mathbf{C}(\rho''_{ABA'B'})\right \|_{\text{tr}}=2\\
    K&\left\|\mathbf{P}(\rho'_{ABA'B'})\right \|_{\text{tr}}=\left\|\mathbf{P}(\rho''_{ABA'B'})\right \|_{\text{tr}}=3.
\end{align}
The norm criteria can therefore be increased by a local operation, even without classical communication. This stems from the fact that these matrix norm methods measure correlations, not entanglement. This can be used as entanglement criteria, as classical correlations cannot surpass a given threshold. But at the same time the norms can be increased by creating additional classical correlations.
Note that this example also holds for less strict criteria than the ones requested by Ref.~\cite{Vidal_2000}.


\section{Proof of the EAM bound}\label{AppEAMproof}

We first show for any operator $\sigma$ and EAM $\{E_i\}_{i=1}^n$ the index of coincidence $I(\sigma)\equiv \sum_{i=1}^n |\tr(E_i \sigma)|^2$ satisfies
\begin{align}\label{EAMcoincidence}
    I(\sigma)\le&\frac{(n-d)\tr(\sigma)^2+d(d-1)\tr(\sigma^\dagger \sigma)}{n(n-1)}.
\end{align}
This extends the result of Rastegin to linear operators and the bound of Eq.~\eqref{rasteginbound} to EAMs.
The general proof follows the same steps as Rastegin~\cite{Rastegin2021} with the following adaptations
\begin{align}
    &a_0=\tr(\sigma)/d\\
    &a_k=\frac{1}{d}\sqrt{\frac{n-1}{d-1}}\sum\limits_{i=1}^n\omega^{-k(i-1)}\tr(E_i \sigma)\\
    &\tr(\sigma^\dagger \sigma)\ge \tr(\sigma)^2/d^2+\sum\limits_{k=1}^{n-1}a_k^*a_k\\
    &\sum\limits_{k=1}^{n-1}a_k^*a_k=\frac{n-1}{d^2(d-1)}(n I(\sigma)-\tr(\sigma)^2).
\end{align}
Eq.~\eqref{EAMcoincidence} follows immediately.

In the same way as for SIC-POVMs, we can restrict our analysis to pure states with Schmidt decomposition $\ket{\psi}=\sum_{s=0}^{r-1} \lambda_s \ket{ss}$. The probabilities are then given by
\begin{align}\nonumber
P_{ab}&=\sum_{s,t=0}^{r-1} \lambda_s \lambda_t \bracket{ss}{E^A_a\otimes E^B_b}{tt}\\
&=\sum_{s=0}^{r-1}\lambda_s^2 D_s^{a,b} +\sum_{s\neq t} \lambda_s \lambda_tO_{s,t}^{a,b},
\end{align} 
with $D_s^{a,b}=\bracket{ss}{E^A_a\otimes E^B_b}{ss}$ and $O_{s,t}^{a,b}=\bracket{ss}{E^A_a\otimes E^B_b}{tt}$. A direct calculation gives
\begin{align}
\|D_{s}\|_\text{tr}&=\sqrt{I_A(\ketbra{s})} \sqrt{I_B(\ketbra{s})}\le\frac{(n-d)+d(d-1)}{n(n-1)}\\
\|O_{s,t}\|_\text{tr}&=\sqrt{I_A(\ketbra{s}{t})} \sqrt{I_B(\ketbra{s}{t})}\le\frac{d(d-1)}{n(n-1)}
\end{align}
Now we can write
\begin{align}
    \|\mathbf{P}_n\|&\le\sum\limits_{s=1}^r\lambda_s^2 \|D_s\|+\sum\limits_{s\neq s}\lambda_s\lambda_t \|O_{s,t}\|\\
    &\le\frac{(n-d)+d(d-1)}{n(n-1)}\sum\limits_{s=1}^r\lambda_s^2+\frac{d(d-1)}{n(n-1)}\sum\limits_{s\neq t}\lambda_s\lambda_t\\
    &=\frac{n-d}{n(n-1)}\sum\limits_{s=1}^r\lambda_s^2+\frac{d(d-1)}{n(n-1)}\sum\limits_{s,t=1}^r\lambda_s\lambda_t\\
    &\le\frac{n-d+d(d-1)r}{n(n-1)}.
\end{align}

By convexity the same result holds for all mixed states of Schmidt number $r$.

\end{document}